\documentstyle[12pt]{article}
\renewcommand \baselinestretch{1.4}

\begin{document}

\def\beq{\begin{equation}}
\def\eeq{\end{equation}}
\def\bce{\begin{center}}
\def\ece{\end{center}}
\def\bea{\begin{eqnarray}}
\def\eea{\end{eqnarray}}
\def\ben{\begin{enumerate}}
\def\een{\end{enumerate}}
\def\ul{\underline}
\def\ni{\noindent}
\def\nn{\nonumber}
\def\bs{\bigskip}
\def\ms{\medskip}
\def\wt{\widetilde}
\def\brr{\begin{array}}
\def\err{\end{array}}

\hfill UB-ECM-PF 94/9

\hfill March, 1994

\vspace*{3mm}

\begin{center}

{\LARGE \bf
Renormalization-group improved effective potential for
finite grand unified theories in curved spacetime}

\vspace{4mm}

\renewcommand\baselinestretch{0.8}
\medskip

{\sc E. Elizalde}
\footnote{E-mail: eli@zeta.ecm.ub.es, eli@ebubecm1.bitnet} \\
Center for Advanced Studies, C.S.I.C., Cam\'{\i} de Santa B\`arbara, 17300
Blanes, \\
and Department E.C.M. and I.F.A.E., Faculty of Physics,
University of  Barcelona, \\ Diagonal 647, 08028 Barcelona, Catalonia,
Spain \\
 and \\
{\sc S.D. Odintsov}\footnote{E-mail: odintsov@ebubecm1.bitnet.
Also at the Department E.C.M., Faculty of Physics,
University of  Barcelona, Diagonal 647, 08028 Barcelona,
Spain.} \\
Tomsk Pedagogical Institute, 634041 Tomsk, Russia. \\

\renewcommand\baselinestretch{1.4}

\vspace{5mm}

{\bf Abstract}

\end{center}

The renormalization-group improved effective potential ---to
leading-log and in the linear curvature approximation--- is
constructed
for ``finite'' theories  in curved spacetime. It is not trivial
and displays
a quite interesting, exponential-like structure ---in contrast
with the
case of flat spacetime where it coincides with the classical
potential.
Several possible cosmological applications, as curvature-induced
phase transitions and modifications of the values of the
gravitational and cosmological
constants, are briefly discussed.

\vspace{4mm}

\newpage

\ni{\bf 1. Introduction}. It is a well-known fact that supersymmetry
(for
an introduction, see \cite{1}) leads to very interesting
consequences. One of the most remarkable is the existence of finite
theories
(see, for instance, \cite{2}-\cite{4} and references therein).
To be
sure, supersymmetric theories are not the only ones which can be
finite
at the one-loop or two-loop level (see \cite{4,6}),
however, up
to now only certain supersymmetric theories have been proven to
be finite to {\it all} orders of perturbation theory
(N=4 super Yang-Mills
theory \cite{5} is the best known example).

Different grand unified theories (GUTs) have been proposed which
turn out to yield finite models. Some of them,
as for instance the finite
supersymmetric SU(5) GUT \cite{7}, may lead to quite reasonable
phenomenological consequences and deserves careful attention as
a realistic model of grand unification. Moreover, some
generalizations of the concept of finite theory ---as the
asymptotically finite GUTs--- have
been proposed \cite{8}. In such theories the zero charge problem
is
absent, both in the UF and in the IR limit, since in these limits
the effective coupling constants tend to some constant values
(corresponding to a finite phase).

When we consider a massless finite or massless asymptotically
finite GUT there is no much sense in discussing quantum
corrections
to the classical potential in such a theory, since either they
are simply absent or are highly suppressed asymptotically.
However, when studying
 finite theories not in flat but in curved spacetime \cite{9}
(for a general review, see \cite{10}) the situation changes
drastically.
There appears a non-trivial effective coupling constant $\xi (t)$
corresponding to the scalar-gravitational interaction which
becomes
important in different respects.

The purpose of this letter is to discuss the issue of the
effective
potential for finite (or asymptotically finite) theories in curved
spacetime. More precisely,
we will construct the renormalization-group (RG) improved
effective
potential ---in the linear curvature and leading-log
approximation--- for finite theories in curved spacetime. It will
be shown that this potential has an interesting and unusual
structure (as
compared whith that for the standard theories). Among the possible
cosmological applications of the model, curvature-induced
phase transitions will be investigated.
Some speculations about the associated variations of the
cosmological and gravitational constants are made at the end.
\ms

\ni{\bf 2. The RG improved effective potential}.
Let us start from a certain multiplicatively-renormalizable
massless
theory in curved spacetime. The general form for the Lagrangian
of such
kind of theory is
\beq
L = \sqrt{-g} \left( L_m + L_{ext} + \xi R \varphi^2 \right),
\label{1}
\eeq
where $L_m$ corresponds to some GUT which includes multiplets of
spinors
$\psi$ and scalars $\varphi$, and gauge fields $A_\mu$, $\xi$ is
the scalar-gravitational coupling constant, and
\beq
L_{ext} = aR^2 + bG + c C_{\mu\nu\alpha\beta}^2 + d \Box R.
\eeq
Here $G$ is the Gauss-Bonnet invariant, $C_{\mu\nu\alpha\beta}$
the Weyl
tensor and $R$ the curvature. Notice that the necessity to
introduce
$L_{ext}$ and $\xi R \varphi^2$ is demanded by renormalizability
in curved spacetime (see \cite{10} for an introduction).

We will consider the situation when $L_m$ corresponds to a finite
(at least to the one-loop approximation) theory in flat space.
Of course, in curved space such a theory is not finite, generally
speaking, due to the extra terms in (\ref{1}) (namely the vacuum
energy and the non-minimal
scalar-gravitational interaction term). This is why we
prefer to call such a theory ``finite'', thus remarking that it is
actually finite in flat space only. Finiteness means that the
effective coupling constants of the theory have the following form
\beq
g^2(t)= g^2, \ \ \ \ \ h^2(t)= \kappa_1 g^2, \ \ \ \ \
\lambda (t)= \kappa_2 g^2,
\label{3}
\eeq
where $g^2$, $h^2$ and $\lambda$ are respectively
the gauge, Yukawa and scalar couplings, $\kappa_1$ and
$\kappa_2$ are some constants, and $g^2 <<1$. Now we are going to
study the RG improved effective potential \cite{11} in such a theory
(for
a very recent discussion in the case of the standard model, together
with a list of references, see \cite{12}). In flat space such
potential is given simply by a classical potential because there
are no one-loop quantum corrections. In curved spacetime the
situation changes a lot. For simplicity, let the classical
potential contain only one scalar multiplet, i.e.
\beq
V_{cl} = a\lambda \Phi^4 -b\xi R \Phi^2,
\eeq
where $a$ and $b$ are some positive constants. Then, by standard
methods one can show that the RG improved effective potential in
curved spacetime is given by \cite{14}
\beq
V = a\lambda (t) f^4(t) \varphi^4 -b\xi (t) f^2(t) R \varphi^2,
\label{5}
\eeq
where $t= (1/2) \ln (\varphi^2 /\mu^2)$, and we have
\bea
&& \dot{\wt{g}} (t) = \bar{\beta}_{\wt{g}} (t), \ \wt{g} (0) =\wt{g};
\ \ \ \dot{\alpha} (t) = \bar{\delta} (t), \ \alpha (0) = \alpha;
\ \ \ \dot{\xi} (t) = \bar{\beta}_\xi (t), \ \xi (0) =\xi; \nn \\
&& f(t) = \exp \left[ - \int_0^t dt' \, \bar{\gamma} \left( \wt{g}
(t'), \alpha (t') \right) \right],
\label{6}
\eea
and $(\bar{\beta}_{\wt{g}}, \bar{\delta}, \bar{\beta}_\xi,
\bar{\gamma} ) = (\beta_g, \delta, \beta_\xi, \gamma ) /(1+ \gamma )$,
$\wt{g} \equiv \{ g, h^2, \lambda \}$, $\alpha$ is the set of gauge
parameters.
For asymptotically free, as well as for non asymptotically free
theories (such as the $\lambda \varphi^4$-theory) the RG improved
effective potential (\ref{5}) has been investigated in \cite{14}
already. Notice that this potential is obtained in the linear
curvature approximation, which is good enough for GUTs
corresponding to the curved spacetime of the early universe \cite{13}.

In ``finite'' theories, the effective coupling constants having
analogues in flat spacetime are given by Eq. (\ref{3}). The
general structure of the one-loop effective coupling constant
$\xi (t)$ for ``finite'' theories in curved spacetime has been
already obtained in Ref. \cite{9} as the solution of the RG
equation
$\dot{\xi} (t) = \beta_\xi (t)$, that is
\beq
\dot{\xi} (t) = \left( \xi (t) - \frac{1}{6} \right) Cg^2,
\label{7}
\eeq
where the constant $C$ is fixed by specific features of the
theory under discussion (it can be equal to zero for some
theories). For $C\neq 0$, we have
\beq
\xi (t) = \frac{1}{6} + \left( \xi - \frac{1}{6} \right) \exp
(Cg^2t).
\label{8}
\eeq

In particular, for the SU(2) finite gauge model of Ref. \cite{4},
we have $C=6$ or $C\simeq 28$ \cite{9}. Hence (non-asymptotical
conformal invariance), in such theories we have $|\xi (t)| \rightarrow
\infty$
in the UF limit ($t\rightarrow \infty$). In the models
which have $C<0$ (examples of this kind of models can be found
in
\cite{15,6}), one obtains $\xi (t) \rightarrow 1/6$ (asymptotical
conformal invariance).

Notice that our theory is defined not only in the UF limit but
also has good behaviour in the  IR limit (see (\ref{3})). In the
IR limit the behaviour of $\xi (t)$ is reversed. In particular,
for $C>0$ one has  $\xi (t) \rightarrow 1/6$ as  $t \rightarrow
-\infty $ (asymptotical confomal invariance), with independence
of the choice of initial value. It is interesting to observe that
there exist models of inflationary universes (the variable
Planck-mass models) for which the typical value of $\xi$ is very
high, $|\xi| \sim 10^4$ \cite{18} (for an introduction to the
inflationary universe, see for instance \cite{19}). The
exponential running of $\xi (t)$ in ``finite'' GUTs (for models
with $C<0$) provides a very natural way to obtain a very large
$\xi$ in the early universe and, at the same time, a $\xi$ close
to the conformal value at present energies. Hence, ``finite'' GUTs
may provide a reasonable justification for the presence of such
a large $\xi$ in the inflationary universe models of \cite{18}.

Now, in order to obtain the RG improved effective potential we
should know the $\gamma$-function of the scalar field in
(\ref{6}). At the one-loop level, $\gamma \sim a_1 g^2 +a_2 h^2$,
where $a_1$ and $a_2$ are some constants whose value depend on
the gauge choice and other features of the theory. As $h^2 =
\kappa_1 g^2$, it turns out that $\gamma \sim (a_1 +\kappa_1
a_2)g^2$. Through the choice of gauge parameter, one can obtain
different values for $\gamma$. As we restrict ourselves to finite
theories, it seems rather reasonable to try to obtain as much
finiteness in our theory as possible. Thus, we will choose the
gauge in which the one-loop $\gamma$-function is equal to zero.
This choice is always possible, moreover in supersymmetric finite
theories it appears in a very natural way (specially if the
superfield technique is used). Having done all this, it turns out
that the RG improved effective potential (in the linear-curvature
and leading-log approximation) for a ``finite'' theory in curved
spacetime is given by
\beq
V= a\kappa_1 g^2 \varphi^4 - b \xi (t) R \varphi^2.
\label{9}
\eeq
It is remarkable that this expression is valid for any $t$ (namely
it is
not restricted by perturbation theory), contrary to what happens
in the standard situation \cite{11}. One can consider its UF or
IR limits (or regions close to those limits) at the same time.

If we choose to keep the gauge arbitrary, and do not demand that
$\gamma$ vanishes, then we get
\beq
V= a\kappa_1 g^2 f^4(t) \varphi^4 - b \xi (t)f^2(t) R \varphi^2,
\label{10}
\eeq
and $\gamma = C_1 g^2$, $C_1$ being some constant which depends
on the gauge parameter and on the features of the theory, $f(t)=
\exp (-C_1 g^2 t)$, and $\xi (t)$ is given by (\ref{8}) (since
the corrections comming from the non-zero $\gamma$ in the
equation $\dot{\xi} (t) = \beta_\xi /(1+\gamma )$ are of two-loop
order).

Finally, for asymptotically finite theories \cite{8} one can get
\beq
V= a\lambda (t) f^4(t) \varphi^4 -b \xi (t) f^2(t) R\varphi^2,
\eeq
where we know only the asymptotic values of $\lambda (t)$, $f(t)$
and $\xi (t)$ in the UF or IR limit, but ignore the
behaviour of these effective couplings in the intermediate
region. Recently, asymptotically finite theories have been used
in \cite{16} as the basis of a possible mechanism for quantum
screening of the cosmological constant (this will be discussed
below).

Hence, we have obtained a non-trivial RG improved effective
potential for ``finite'' theories in curved spacetime. The most
interesting property of this potential is the fact that it is
valid, generally speaking, for {\it any} value of $t$ (provided,
of course, that we are inside a region where perturbation theory
has sense).
\ms

\ni{\bf 3. Curvature-induced phase transitions}. It is common
belief that the early universe has experienced one or several
phase transitions. Such phase transitions could have been caused
by modifications of some external parameters of the theory (as
temperature, pressure or an external gravitational field). The
inflationary phase of the early universe could probably have
resulted from a phase transition of Coleman-Weinberg type. We
will here discuss the possibility of a curvature-induced phase
transition in a ``finite'' theory in curved spacetime, thus
assuming as starting point that, in fact, the early universe can
be described by a convenient ``finite'' GUT. As before, we shall
work with the RG improved effective potential. The first
realization of a gravitational phase transition, for the case of
scalar QED in the De Sitter space, has been presented in Ref.
\cite{13} (for a review and a list of references, see \cite{10}).

We will be interested in first order phase transitions, where the
order parameter $\varphi$ experiences a quick change for some
critical value, $R_c$, of the curvature (see, for example,
\cite{10}). As usually \cite{13,14}, it is convenient to
 introduce the
dimensionless parameters $x= \varphi^2/\mu^2$,  $y= |R|/\mu^2$
and $\epsilon = $ sgn $R$, in terms of which the RG improved
effective potential is given by (for simplicity, we consider the
potential (\ref{9}))
\beq
\frac{V}{\mu^4}=  a \kappa_1 g^2 x^2-b \epsilon x y \left[
\frac{1}{6} + \left( \xi  - \frac{1}{6} \right) x^{Cg^2/2}
\right].
\eeq
The critical parameters corresponding to the first-order phase
transition are defined by the standard conditions:
\beq
V(x_c,y_c) =  0, \ \ \ \ \ \  \left. \frac{\partial V}{\partial
x} \right|_{x_c,y_c} =  0, \ \ \ \ \ \    \left. \frac{\partial^2
V}{\partial x^2} \right|_{x_c,y_c} >  0.
\label{13}
\eeq
The analysis of curvature-induced phase transitions, both for
asymptotically free and for asymptotically non-free theories
---using the corresponding RG improved effective potential---
 has been carried out in Ref. \cite{14}.

For ``finite'' theories with the potential (\ref{9}), the
analysis of Eqs. (\ref{13}) yields the following critical values
for the phase transition
\beq
x_c = \left[  ( 1- 6 \xi ) \left( 1 -
\frac{Cg^2}{2} \right)\right]^{-2/Cg^2}, \ \ \ \ \ \epsilon y_c
= \frac{12a\kappa_1}{bC} \left( -1 + \frac{Cg^2}{2} \right) x_c.
\eeq
In addition, we must restrict $a\kappa_1 g^2 (1-Cg^2/2) >0$
(this is the third condition in (\ref{13})) and also
$|12 a\kappa_1  (1- Cg^2/2)/(bC)| <1$ (otherwise the
linear curvature approximation is broken).

Let us now produce some simple estimation of a possible result.
Guided by
the explicit values corresponding to the SU(2) finite gauge
theories \cite{4,9,15}, for negative $C$ a reasonable
choice of parameters satisfying Eqs. (\ref{13}) may be the
 following:
\beq
b= \frac{1}{2}, \ \ \ a= \frac{1}{4}, \ \ \ \kappa_1
= 0.1, \ \ \ C= -10.
 \eeq
Choosing moreover a small coupling constant $g^2= 0.05$ and
an initial value $\xi = 0.15$, one  gets
\beq
x_c = 2 \cdot 10^{-4}, \ \ \ \ \ \ |y_c| = 1.5 \cdot 10^{-5}.
\label{16}
\eeq
This estimation is already very much close to the standard value of
the curvature in the GUT epoch, namely
\beq
10^{-7} \leq |y| \leq 10^{-5}.
\eeq
We have thus shown, in principle, the possibility of a
gravitational phase transition for ``finite'' GUTs in curved
spacetime. In fact this simple calculation hints very clearly
 towards the possibility of
emergence of some kind of inflationary stage in the evolution of
the early universe, based on the occurrence of a gravitational
phase transition that, as we see, can appear quite naturally
when we use a ``finite'' GUT to describe the early universe.
Due to its
simplicity, this is a rather remarkable result that deserves
further study.
\ms

\ni{\bf 4. Induced cosmological and gravitational constants}. It is
a matter of fact that (at least in principle) in the assumed
situation that the early universe can be described by some
finite or asymptotically finite GUT, one can estimate the induced
values of the cosmological and gravitational constants ---or, more
precisely, of their variations--- by using the RG improved
effective potential. Consider the RG improved effective potential
under its most general form (\ref{10}). By making the
identifications
\beq
G^{-1} \sim \xi (t) f^2(t) \varphi^2, \ \ \ \ \ \ \Lambda \sim -
 \frac{g^2f^2(t) \varphi^2}{\xi (t)},
\label{17}
\eeq
one can actually estimate, in principle, the comological
variations of the induced effective cosmological and Newton
constants. In particular, by supposing \cite{16} that a
``finite'' GUT phase of the early universe is appropriate from
some energy scale $\mu_{UV}$ ($t_{UV}$ is large, i.e. the
curvature is strong enough) to another scale  $\mu_{IR}$
($t_{IR}$ is small, i.e. the curvature is very weak), one can
see that the variations of $G$ and $\Lambda$ are given by
\beq
\frac{G^{-1}_{IR}}{G^{-1}_{UV}} \sim \left(
\frac{\mu_{IR}}{\mu_{UV}} \right)^{(C-2C_1)g^2}, \ \ \ \ \ \
\frac{\Lambda_{IR}}{\Lambda_{UV}} \sim \left(
\frac{\mu_{IR}}{\mu_{UV}} \right)^{-(C+2C_1)g^2}.
\eeq
If one wants now to suppress the variation of the Newton
constant, one can choose the gauge (and also, perhaps, the model)
in order to obtain $C=2C_1$. In this case an exponentially large
screening of the cosmological constant appears which takes it
from a very large value in the early universe to the extremely
small one that it has at present energies. This simple mechanism,
called RG screening of the cosmological constant, has been proposed
in Ref. \cite{16}. Unfortunately, the special choice that one has
to do of the $\gamma$-function (or, in other words, of the constant
 $C_1$) is
a kind of fine-tuning of the model. This feature, albeit common
to almost any solution that has been proposed of the cosmological
constant problem, is not very appealing.

In principle, the variations of $G$ and $\Lambda$ in ``finite''
GUTs depend very much  on the choice of the model and gauge. In
particular, for the `natural' choice $C_1=0$ we get an
exponential screening of  $\Lambda$ but, at the same time, an
exponential growth of the Newton constant. Such an outcome does
not look very realistic either.
\ms

\ni{\bf 5. Conclusion}. In this note we have studied in some
detail the RG improved effective potential corresponding to
massless  ``finite'' GUTs in curved spacetime. Some cosmological
speculations involving potentials of this kind have been
analyzed. However, there is still a number of related questions
that are left for further study. In particular, the
important question of how a non-zero mass would influence the
results of the RG potential. It is known that the answer to this
question is not easy \cite{12,20}, owing to the fact that usually
one has simultaneously several effective masses in the theory and
also because of the running of the vacuum energy, already in flat
space. In curved space the situation is even more complicated and
one is led to study the RG improved effective Lagrangian actually
\cite{17}.

Another question concerns the generality of such a type of RG
improved potential and if the results that we have obtained
will be extrapolable (at least qualitatively) to other
theories. To answer this, let us consider the $d=3$ abelian
Chern-Simons gauge
theory (for an inntroduction, see \cite{21,22}) interacting with
matter. The interest of such kind of theories is mainly motivated
by the possibility of their use as models of high-temperature
superconductivity. The property of being finite is a very
natural one in the case of an abelian Chern-Simons gauge
theory \cite{23}.
Starting from a Chern-Simons theory with matter in a $d=3$ curved
spacetime \cite{24}, one can construct the corresponding RG
improved effective potential, what gives a similar result to that
of Sect. 3, namely
\beq
V=h(t) \Phi^6 (t) - \xi (t) R \Phi^2 (t),
\eeq
where $\Phi^2 = \Phi^* \Phi$, $h$ is the scalar coupling
constant, $h(t) = \kappa_1 e^4$ (in the regime where the theory
is finite), $e^2$ is the electromagnetic coupling constant, and
$e^2(t) = e^2$ \cite{23}. The effective field is given by $\Phi
(t) = \exp (-C_1 e^4 t) \Phi$ and $\xi (t) = 1/8 + (\xi - 1/8)
\exp (C_1e^4t)$ \cite{24}. Hence, in the case of an abelian
Chern-Simons theory with matter in curved spacetime, the RG
improved effective potential has the same functional form as in
the case of a ``finite'' GUT, and may also lead to $d=3$
curvature-induced phase transitions.
\vspace{5mm}

\noindent{\large \bf Acknowledgments}

SDO would like to thank the members of the Dept. ECM, Barcelona
University, for very kind hospitality.
This work has been supported by DGICYT (Spain), project no.
PB90-0022, and by CIRIT (Generalitat de Catalunya).


\newpage


\begin{thebibliography}{99}

\bibitem{1} P. West, {\it Introduction to supersymmetry and
supergravity} (World Scientific, Singapore, 1986).

\bibitem{2} D.R.T. Jones and L. Mezinescu, Phys. Lett. {\bf B136}
(1984) 243; {\bf B138} (1984) 293; A. Parkes and P. West, Phys.
Lett. {\bf B138} (1984) 99; {\bf B127} (1983) 353; J.-M. Frere,
L. Mezinescu and J.-P. Yao, Phys. Rev. {\bf D29} (1984) 1196; S.
Hamidi and J.H. Schwarz, Phys. Lett. {\bf B147} (1984) 301;
S. Rajpoot and J.G. Taylor, Phys. Lett. {\bf B147} (1984) 91; J.-P.
Derendinger, S. Ferrara and A. Masiero, Phys. Lett. {\bf B143}
(1984) 133; S. Kalara, D. Chang, R.N. Mohapatra and A.
Gangopadhyay, Phys. Lett. {\bf B145} (1984) 323.

\bibitem{3} P. Fayet, Phys. Lett. {\bf B153} (1985) 397; J.E.
Bjorkman, D.R.T. Jones and S. Raby, Nucl. Phys. {\bf B259} (1985)
503; O. Piguet and K. Sibold, Phys. Lett. {\bf B177} (1986) 373.

\bibitem{4} M. B\"{o}hm and A. Denner, Nucl. Phys. {\bf B282} (1987)
206.

\bibitem{6} I.L. Shapiro and E.G. Yagunov, Int. J. Mod. Phys.
{\bf A8} (1993) 1787; W. Lucha and F.F. Sch\"{o}berl, preprint
UWThPh-1993--56 (1993).

\bibitem{5} P.S. Howe, K.S. Stelle and P.K. Townsend, Nucl. Phys.
{\bf B214} (1983) 519; {\bf B236} (1984) 125.

\bibitem{7} S.D. Odintsov and I.L. Shapiro, JETP Lett. {\bf 49}
(1989) 125; Mod. Phys. Lett. {\bf A4} (1989) 1479.

\bibitem{8} A.V. Ermushev, D.I. Kazakov and O.V. Tarasov,
 Nucl. Phys. {\bf B281} (1987) 72; D. Kapetanakis, M. Mondrag\'on
and G. Zoupanos, Z. Phys. {\bf C60} (1993) 181.

\bibitem{9} I.L. Buchbinder, S.D. Odintsov and I.M. Lichtzier,
Class. Quant. Grav. {\bf 6} (1989) 6055.

\bibitem{10} I.L. Buchbinder, S.D. Odintsov and I.L. Shapiro,
{\sl Effective Action in Quantum Gravity} (IOP Publishing,
Bristol and Philadelphia, 1992).

\bibitem{11} S. Coleman  and E. Weinberg,  Phys. Rev. {\bf D7}
(1973) 888.

\bibitem{12}  C. Ford, D.R.T. Jones, P.W. Stephenson and
M.B. Einhorn, Nucl. Phys. {\bf B395} (1993) 17.

\bibitem{14} E. Elizalde and S.D. Odintsov, Phys. Lett. {\bf
B303} (1993) 240.

\bibitem{13} G.M. Shore, Ann. Phys. (NY) {\bf 128} (1980) 376.

\bibitem{15} S.D. Odintsov and F.Sh. Zaripov, Mod. Phys. Lett.
{\bf A4} (1989) 1955.

\bibitem{18} D. Salopek, J.R. Bond and J.M. Bardeen, Phys. Rev.
{\bf D40} (1989) 1753.

\bibitem{19} E.W. Kolb and M.S. Turner, {\it The Early
Universe} (Addison-Wesley, Reading, MA, 1990).

\bibitem{16} I.L. Shapiro, preprint HUPD-9405 (1994).

\bibitem{20} M. Bando, T. Kugo, N. Maekawa and H. Nakano,
Prog. Theor. Phys. {\bf 90} (1993) 405.

\bibitem{17} E. Elizalde and S.D. Odintsov, Phys. Lett. {\bf
B321} (1994) 199.

\bibitem{21} S. Deser, R. Jackiw and  S. Templeton, Ann. Phys.
(NY) {\bf 140} (1982) 372.

\bibitem{22} A. Niemi and G. Semenoff, Phys. Rep. {\bf 135}
(1987) 3.

\bibitem{23} L.V. Avdeev, D.I. Kazakov and G.V. Grigoryev,
preprint CERN-6091 (1991).

\bibitem{24} S.D. Odintsov, Z. Phys. {\bf C54} (1992) 527.

\end{thebibliography}
\end{document}